\documentstyle[pra,aps]{revtex}
\begin{document}
\draft
\title{Quantum radiation in a plane cavity with moving mirrors }
\author{D. F. Mundarain$^1$ and P. A. Maia Neto$^2$}
\address{$^1${\it Facultad de F\'{\i}sica, Pontificia Universidad Cat\'{o}lica, }%
Casilla 306, Santiago 22, Chile\\
$^2${\it Instituto de F\'{\i}sica, UFRJ, Caixa Postal 68528, 21945-970 Rio }%
de Janeiro, Brazil}
\date{\today}
\maketitle

\begin{abstract}
We consider the electromagnetic vacuum field inside a perfect plane cavity
with moving mirrors, in the nonrelativistic approximation. We show that low
frequency photons are generated in pairs that satisfy simple properties
associated to the plane geometry. We calculate the photon generation rates
for each polarization as functions of the mechanical frequency by two
independent methods: on one hand from the analysis of the boundary
conditions for moving mirrors and with the aid of Green functions; and on
the other hand by an effective Hamiltonian approach. The angular and
frequency spectra are discrete, and emission rates for each allowed angular
direction are obtained. We discuss the dependence of the generation rates on
the cavity length and show that the effect is enhanced for short cavity
lengths. We also compute the dissipative force on the moving mirrors and
show that it is related to the total radiated energy as predicted by energy
conservation.
\end{abstract}


\section{Introduction}

In the presence of moving boundaries, the vacuum state of the
electromagnetic field may not be stable, then resulting in the generation of
photons. This purely quantum effect, which has been known either as
dynamical Casimir effect~\cite{schwinger} or as motion~\cite{astrid} or
mirror~\cite{barton} induced radiation is, like the usual Casimir effect for
standing mirrors, a striking illustration of the physical reality of the
quantum vacuum field. Moreover, it may also be understood as a mechanical
effect of the vacuum field. In fact, energy conservation entails that the
radiation effect must be accompanied by a radiation reaction force that
works against the motion of the mirror~\cite{barton}--\cite{jr1}, and which
is connected to the fluctuations of the usual (static) Casimir force by the
fluctuation--dissipation theorem~\cite{fusao}--\cite{motion}.

Several theoretical models have been analyzed. In the one--dimensional
approximation (1D), only one direction of propagation is taken into account~%
\cite{moore}. The quantum radiation generated inside a 1D cavity with moving
mirrors was calculated in Refs.~\cite{Law} and \cite{dodo} in the particular
case where the mechanical frequency satisfies a resonant condition for
generation of photons in the lowest order cavity field modes, whereas Ref.~%
\cite{astrid} considered a 1D cavity with partially--transmitting mirrors
and with no particular assumption about resonance, thereby allowing for a
full analysis of the spectrum of the radiation in a more general case.

A few three--dimensional (3D) models have been recently analyzed in the
literature, including moving dielectric half-spaces \cite{barside2}--\cite
{north}, and rotating~\cite{barside} or collapsing dielectric spheres~\cite
{eberlein}, the latter as a model for sonoluminescence. On the other hand,
3D results  for the related problem of photon generation in a medium with
time--dependent material coefficients $\epsilon$ and $\mu$ have been known
for nearly ten years~\cite{bb}. Perhaps the simplest 3D illustration of
motion induced radiation is to consider a single perfectly--reflecting plane
mirror moving in free space. In the perturbative regime, which is associated
to the nonrelativistic limit, it is possible to derive simple results for
the spectra of radiation \cite{pa4}, which display interesting
polarization--dependent features connected to the angular distribution of
the emitted photons. In this paper, we extend the method developed in Ref.~%
\cite{pa4} to analyze the radiation emitted when two parallel plane 
perfectly reflecting mirrors,
initially a distance $L$ apart, oscillate along the direction perpendicular
to their surfaces, and according to a pre--defined law imposed by some
external apparatus. Such geometry constitutes the simplest example, from a
theoretical point--of--view, of a 3D cavity of length $L.$ As compared to
the previous single--mirror case, we show that the orders of magnitude for
the radiation rates generated in the plane cavity may be several orders of
magnitude larger, provided that $L$ is small enough.

The paper is organized as follows. In Sec.~II, we calculate the photon
numbers generated inside the cavity starting from the boundary conditions of
a moving perfectly reflecting mirror. The method is based on the
nonrelativistic and long--wavelength approximations, which are closely
connected in the context considered here~\cite{pa1}. In Sec.~III we present
an alternative derivation of the results already found in Sec.~II, now
employing usual time--dependent perturbation theory for an effective
Hamiltonian which incorporates the motion effect in terms of a coupling via
radiation pressure. This heuristic approach is considerably simpler than the
previous one, since it circumvents the analysis of the moving boundaries.
Furthermore, it explicitly unveils the two--photon nature of the photon
emission process, and allows for the computation of the dissipative
component of the radiation pressure force on the moving mirrors. In Sec.~IV,
we consider an specific example of motion in order to isolate the effect of
a single mechanical frequency $\omega _0.$ We show that the photon numbers
obtained by two independent methods in Secs.~II and III grow linearly in
time, allowing us to define photon production rates, whose behavior as
functions of the dimensionless parameter $\omega _0L/\pi c$ is examined in
detail. Sec.~V contains the concluding remarks.

\section{Boundary conditions and intracavity quantum radiation}

For the sake of clarity we first assume that one of the mirrors is at rest.
The results in the more general case where both mirrors are set to move is a
simple generalization to be presented later. The moving mirror
oscillates along the direction perpendicular to its surface
($x$ direction), around the position $x=0, $ its
instantaneous position being given by the equation $x=\delta \!q(t).$

We decompose the electromagnetic fields into their components corresponding
to the electric field parallel (TM) or perpendicular (TE) to the plane of
incidence. For each polarization it is possible to define a vector potential
through the equations: 
\begin{eqnarray}
{\bf E}^{{\rm \scriptscriptstyle (TE)}}=-\partial _t{\bf A}^{{\rm %
\scriptscriptstyle (TE)}};{\bf B}^{{\rm \scriptscriptstyle (TE)}}=\nabla
\times {\bf A}^{{\rm \scriptscriptstyle (TE)}}
\end{eqnarray}
and 
\begin{eqnarray}
{\bf E}^{{\rm \scriptscriptstyle (TM)}}=\nabla \times 
\mbox{\boldmath$\cal
A$}^{{\rm \scriptscriptstyle (TM)}};{\bf B}^{{\rm \scriptscriptstyle (TM)}%
}=\partial _t\mbox{\boldmath$\cal
A$}^{{\rm \scriptscriptstyle (TM)}}.  \label{defitm}
\end{eqnarray}
The units are MKS with $c=1$ and $\varepsilon _0=1$. 
The potentials satisfy the Gauge equations 
\begin{equation}
\nabla \cdot {\bf A}^{{\rm \scriptscriptstyle (TE)}}=\nabla \cdot %
\mbox{\boldmath$\cal A$}^{{\rm \scriptscriptstyle (TM)}}=0.
\end{equation}
As shown in Appendix A, the boundary conditions for a perfectly reflecting
moving mirror are very simple when written in terms of ${\bf A}^{{\rm %
\scriptscriptstyle (TE)}}$ and $\mbox{\boldmath$\cal A$}^{{\rm %
\scriptscriptstyle (TM)}},$ due essentially to the fact that they are both
orthogonal to the direction of motion. We find 
\begin{equation}
{\bf A}^{{\rm \scriptscriptstyle (TE)}}(x=\delta \!q(t),{\bf r}_{\parallel
},t)=0  \label{bc1}
\end{equation}
and 
\begin{equation}
(\partial _x+\delta \!\dot{q}(t)\partial _t)\mbox{\boldmath$\cal A$}^{{\rm %
\scriptscriptstyle (TM)}}(x=\delta \!q(t),{\bf r}_{\parallel },t)=0,
\label{bc2}
\end{equation}
where ${\bf r}_{\parallel }=y\hat{y}+z\hat{z}.$ Furthermore, the fields
satisfy the usual homogeneous Dirichlet and Neumann boundary conditions on
the second mirror, which is at rest at $x=L:$ 
\begin{eqnarray}
{\bf A}^{{\rm \scriptscriptstyle (TE)}}(x=L,{\bf r}_{\parallel
},t)=0;\partial _x\mbox{\boldmath$\cal A$}^{{\rm \scriptscriptstyle (TM)}%
}(x=L,{\bf r}_{\parallel },t)=0.  \label{bc3}
\end{eqnarray}
We want to solve the boundary value problem as defined by Eqs.~(\ref{bc1})--(%
\ref{bc3}) for the fields in the region between the mirrors. The results for
the fields outside the plane cavity are essentially the same as those for a
single moving mirror in vacuum, and hence may be found in Refs.~\cite
{pa1,pa4}. The essential `ansatz' which allows us to employ the long
wavelength approximation to solve the boundary value problem defined by
Eqs.~(\ref{bc1})--(\ref{bc3}) is to assume that a given mechanical frequency 
$\omega _0$ induces the generation of photons only in the spectral range $%
\omega <\omega _0.$ This property is satisfied by the nonrelativistic models
considered previously (see Refs.~\cite{astrid} and \cite{pa4}). Moreover, it
agrees with the intuitive notion that the radiation effect is a nonadiabatic
process, so that high frequency field modes cannot be excited since the
corresponding time scales are shorter than mechanical time scales
(quasi-static limit). More importantly, we show later in this section that
this property is fully satisfied for the model considered here. As for the
connection with the long wavelength approximation, we note that the
amplitude $\delta \!q_0$ of a sinusoidal nonrelativistic motion must satisfy 
$\omega _0\delta \!q_0\ll 1.$ When combined with our ansatz, this condition
leads to $\delta \!q_0\ll \lambda ,$ where $\lambda $ is the wavelength of
the emitted radiation. Actually, we may be slightly more general and
consider any nonrelativistic oscillatory motion around $x=0$ such
that its Fourier components satisfy the above requirements (more 
specifically, we shall consider  a weakly--damped sinusoidal 
nonrelativistic motion in Sec. IV).

Accordingly, we look for perturbative solutions in the form: 
\begin{equation}
{\bf A}^{{\rm \scriptscriptstyle (TE)}}={\bf A}_{{\rm sta}}^{{\rm %
\scriptscriptstyle (TE)}}+\delta \!{\bf A}^{{\rm \scriptscriptstyle (TE)}},
\end{equation}
and 
\begin{equation}
\mbox{\boldmath$\cal A$}^{{\rm \scriptscriptstyle (TM)}}=\mbox{\boldmath$%
\cal A$}_{{\rm sta}}^{{\rm \scriptscriptstyle (TM)}}+\delta \!%
\mbox{\boldmath$\cal A$}^{{\rm \scriptscriptstyle (TM)}}.
\end{equation}
${\bf A}_{{\rm sta}}^{{\rm \scriptscriptstyle (TE)}}$ and $%
\mbox{\boldmath$\cal
A$}_{{\rm sta}}^{{\rm \scriptscriptstyle (TM)}}$ are the fields satisfying
the Dirichlet and Neumann boundary conditions for standing mirrors, whereas $%
\delta \!{\bf A}^{{\rm \scriptscriptstyle (TE)}}$ and $\delta \!%
\mbox{\boldmath$\cal A$}^{{\rm \scriptscriptstyle (TM)}}$ represent the
first--order modifications induced by the motion. As we show below, they are
smaller than the fields for the static configuration by a factor of the
order of $\delta\! q/ \lambda.$  We expand the fields in Eqs.~(\ref{bc1})
and (\ref{bc2}) in Taylor series around $x=0.$ Since the $j$-th spatial
derivative of a monochromatic travelling wave satisfies 
\begin{equation}
| \partial_x^j {\bf A}|\le ({\frac{2\pi}{\lambda}})^j |{\bf A}|,
\end{equation}
we find from Eq.~(\ref{bc1}) that the TE--polarized field $\delta \!{\bf A}^{%
{\rm \scriptscriptstyle (TE)}}$ is given up to first order in $\delta\! q/
\lambda$ by 
\begin{equation}
\delta \!{\bf A}^{{\rm \scriptscriptstyle (TE)}}(x=0,{\bf r}_{\parallel
},t)=-\delta \!q(t)\partial _x{\bf A}_{{\rm sta}}^{{\rm \scriptscriptstyle %
(TE)}}(x=0,{\bf r}_{\parallel },t).  \label{bcte}
\end{equation}
Note that we have neglected the term $\delta \!q(t)\partial _x \delta \!{\bf %
A}^{{\rm \scriptscriptstyle (TE)}}(x=0,{\bf r}_{\parallel },t)$ because, as
shown by the above result, $\delta \!{\bf A}^{{\rm \scriptscriptstyle (TE)}}$
is already of first order in $\delta \!q/\lambda.$ Following the same method
we find the following result for TM polarization: 
\begin{equation}
\partial _x\delta \!\mbox{\boldmath$\cal A$}^{{\rm \scriptscriptstyle (TM)}%
}(x=0,{\bf r}_{\parallel },t)=-(\delta \!q(t)\partial _x^2+\delta \!{\dot{q}%
(t)}\partial _t)\mbox{\boldmath$\cal
A$}_{{\rm sta}}^{{\rm \scriptscriptstyle (TM)}}(x=0,{\bf r}_{\parallel },t) ,
\label{bctm}
\end{equation}
where now we have also neglected terms of the order of $\delta \!q \delta \!%
\dot{q}/\lambda.$ According to our `ansatz', when considering the generation
of photons out of the vacuum field induced by a mechanical frequency $%
\omega_0,$ the relevant wavelengths are larger than $2\pi/ \omega_0,$ and
thus the neglected terms are all of the order of $(\delta\!\dot{q})^2.$ We
have then transformed  the homogeneous boundary conditions for the total
fields at the time dependent position $x=\delta \!q(t)$ given by Eqs.~(\ref
{bc1}) and (\ref{bc2}) into inhomogeneous boundary conditions for $\delta \!%
{\bf A}^{{\rm \scriptscriptstyle (TE)}}$ and $\delta \!\mbox{\boldmath$\cal
A$}^{{\rm \scriptscriptstyle (TM)}}$ at the position $x=0,$ given by Eqs.~(%
\ref{bcte}) and (\ref{bctm}), which may be solved by standard Green function
techniques.

We introduce periodic boundary conditions on the transverse plane $yz$ over
a surface of area $S.$ In the static case, the normal mode decomposition of
the fields in the interval between the mirrors, $0\le x\le L,$ is then
written as follows: 
\begin{equation}
{\bf A}_{{\rm sta}}^{{\rm \scriptscriptstyle (TE)}}({\bf r}%
,t)=i\sum_{\ell=1}^{\infty} \sum_{n}\sqrt{\frac \hbar {\omega _n^\ell SL}}
\sin ({\frac{\ell\pi}{L}} x)e^{i{\bf k}_{\parallel }^n\cdot {\bf r}%
_{\parallel }}e^{-i\omega _n^\ell t}a_{n,\ell}^{{\rm \scriptscriptstyle (TE)}%
} \hat{{\bf \epsilon}}_n+{\rm H.c.}  \label{nmte}
\end{equation}
and for the TM polarization, 
\begin{equation}
\mbox{\boldmath$\cal A$}_{{\rm sta}}^{{\rm \scriptscriptstyle (TM)}}({\bf r}%
,t)=i\sum_{\ell=0}^{\infty}\sum_{n}\sqrt{\frac \hbar {(1+\delta_{\ell 0})
\omega _n^\ell SL}} \cos ({\frac{\ell\pi}{L}} x)e^{i{\bf k}_{\parallel
}^n\cdot {\bf r}_{\parallel }}e^{-i\omega _n^\ell t}a_{n,\ell}^{{\rm %
\scriptscriptstyle (TM)}} \hat{{\bf \epsilon}}_n+{\rm H.c.},  \label{nmtm}
\end{equation}
where 
\begin{equation}
{\bf k}_{\parallel }^n=2\pi (n_y\widehat{{\bf y}}+n_z\widehat{{\bf z}})/ 
\sqrt{S}
\end{equation}
represents the component of the wavevector parallel to the mirrors --- the
shorthand $n=(n_y,n_z)$ representing a pair of integer numbers. Note that
the two potentials describing orthogonal polarizations are written in terms
of the same unit vector 
\begin{equation}
\hat{{\bf \epsilon}}_n=\hat{{\bf x}}\times {\frac{{\bf k}_{\parallel }^n}{%
k_{\parallel }^n}}.
\end{equation}
Throughout the paper, the sum over $n$ --- as in Eqs.~(\ref{nmte}) and (\ref
{nmtm}) --- runs from $n_y=-\infty$ and $n_z=-\infty$ to $n_y=\infty$ and $%
n_z=\infty.$ A given mode with indexes $(n,\ell)$ corresponds to a standing
wave along the $x$-direction with wavevector $k_x^{\ell}=\ell\pi /L$
travelling along a direction parallel to the mirrors with wavevector ${\bf k}%
_{\parallel }^n.$ Its frequency is given by 
\begin{equation}
\omega _n^\ell=\sqrt{\left( \frac{\ell\pi }L\right) ^2+\frac{\left( 2\pi
\right) ^2}S\Bigl[(n_y)^2+(n_z)^2\Bigr]}.
\end{equation}
The bosonic field operators in Eqs.(\ref{nmte}) and (\ref{nmtm}) satisfy the
usual commutation relations 
\begin{equation}
[a_{n,\ell}^{j}, a_{n^{\prime},\ell^{\prime}}^{j^{\prime}}] = 0,
\label{com1}
\end{equation}
and 
\begin{equation}
[a_{n,\ell}^{j}, (a_{n^{\prime},\ell^{\prime}}^{j^{\prime}})^{\dag}] =
\delta_{n,n^{\prime}}\delta_{\ell,\ell^{\prime}}\delta_{j,j^{\prime}},
\label{com2}
\end{equation}
where $j={\rm TE},{\rm TM}$ represents the polarization.

It is convenient to work with a mixed Fourier representation defined as 
\begin{equation}
{\bf A}_n^{{\rm \scriptscriptstyle (TE)}}[x,\omega]=\frac 1S\int dt\int d^2%
{\bf r}_{\parallel } e^{-i{\bf k}_{\parallel }^n\cdot {\bf r}_{\parallel
}}e^{i\omega t}{\bf A} ^{{\rm \scriptscriptstyle (TE)}}(x,{\bf r}_{\parallel
},t)  \label{Fourier}
\end{equation}
with an analogous expression for TM polarization. The Fourier transformed
fields representing the motion induced perturbation satisfy the 1D
Klein-Gordon equation 
\begin{eqnarray}
\left( \partial _x^2+\omega ^2-(k_{\parallel }^n)^2\right) \delta {\bf \label
{difte} A}_n^{{\rm \scriptscriptstyle (TE)}} \lbrack x,\omega \rbrack&=&0, \\
\left( \partial _x^2+\omega ^2-(k_{\parallel }^n)^2\right) \delta %
\mbox{\boldmath$\cal A$}_n^{{\rm \scriptscriptstyle (TM)}} \lbrack x,\omega
\rbrack&=&0,
\end{eqnarray}
and the boundary conditions at $x=0$ and 
$x=L$
given by Eqs.~(\ref{bc3}), (\ref{bcte}) and (\ref
{bctm}).  The resulting boundary value
problem for TE polarization is solved with the aid of the appropriate
Dirichlet Green function: 
\begin{equation}
G_{n\, \omega}^D(x,x^{\prime })=\frac 2L\sum_{\ell=1}^{\infty} \frac {\sin (%
\frac{\ell\pi x}L)\sin (\frac{\ell\pi x^{\prime }}L)} { (\omega\pm
i\epsilon)^2-{\omega_n^{\ell}}^2},  \label{gd}
\end{equation}
where the plus (minus) sign in Eq.~(\ref{gd}) provides the retarded
(advanced) Green function. The fields with TM polarization are obtained from
the Neumann Green function: 
\begin{equation}
G_{n\, \omega}^N(x,x^{\prime })=\frac 2L\sum_{\ell =0}^{\infty} \frac {\cos (%
\frac{\ell\pi x}L)\cos (\frac{\ell\pi x^{\prime }}L) } {(1+\delta_{\ell
0})((\omega\pm i\epsilon)^2-{\omega_n^{\ell}}^2)}.  \label{gn}
\end{equation}

We assume that the mirror moves during a finite time interval, then
returning to its initial position at $x=0.$ As a consequence, we may define
input and output fields, ${{\bf A}_{{\rm in}}}_n^{{\rm \scriptscriptstyle %
(TE)}}$ and ${{\bf A}_{{\rm out}}}_n^{{\rm \scriptscriptstyle (TE)}}$ 
corresponding to the limit values of very small and very large times (and
likewise in the case of TM polarization), which satisfy the boundary
conditions for a mirror at rest at $x=0.$ They are connected by a suitable
combination of retarded (superscript $R$) and advanced (superscript $A$)
Green functions: 
\begin{eqnarray}
{{\bf A}_{{\rm out}}}_n^{{\rm \scriptscriptstyle (TE)}}[x,\omega ] &=&{{\bf A%
}_{{\rm in}}}_n^{{\rm \scriptscriptstyle (TE)}} [x,\omega ]+\delta {\bf A}%
_n^{{\rm \scriptscriptstyle (TE)}}[x^{\prime }=0,\omega ]\Bigr[\partial
_{x^{\prime }}G_{n\,\omega }^{D,R}(x,x^{\prime }=0)  \nonumber \\
&&-\partial _{x^{\prime }}G_{n\,\omega }^{D,A}(x,x^{\prime }=0)\Bigl].
\label{doug1}
\end{eqnarray}
The TM output field ${\mbox{\boldmath$\cal A$}_{{\rm out}}}_n^{{\rm %
\scriptscriptstyle (TM)}}$ is related to the TM input field ${%
\mbox{\boldmath$\cal A$}_{{\rm in}}}_n^{{\rm \scriptscriptstyle (TM)}}$ by a
similar expression: 
\begin{eqnarray}
{\mbox{\boldmath$\cal A$}_{{\rm out}}}_n^{{\rm \scriptscriptstyle (TM)}%
}[x,\omega ] &=&{\mbox{\boldmath$\cal A$}_{{\rm in}}}_n^{{\rm %
\scriptscriptstyle (TM)}}[x,\omega ]-\partial _{x^{\prime }}\delta %
\mbox{\boldmath$\cal A$}_n^{{\rm \scriptscriptstyle (TE)}}[x^{\prime
}=0,\omega ]\Bigl[G_{n\,\omega }^{N,R}(x,x^{\prime }=0)  \nonumber \\
&&-G_{n\,\omega }^{N,A}(x,x^{\prime }=0)\Bigr].  \label{doug2}
\end{eqnarray}
From Eqs.~(\ref{gd}) and (\ref{gn}) we find 
\begin{eqnarray}
\partial _{x^{\prime }}G_{n\,\omega }^{D,R}(x,x^{\prime } =0)-\partial
_{x^{\prime }}G_{n\,\omega }^{D,A}(x,x^{\prime }=0)&
=-{\frac{2\pi ^2i}{L^2}}%
\sum_{\ell =1}^\infty {\frac \ell {\omega _n^\ell }}\sin (\frac{\ell \pi x}L)
\nonumber \\
&\times \left( \delta (\omega -\omega _n^\ell )-\delta (\omega +\omega
_n^\ell )\right) ,  \label{gdd}
\end{eqnarray}
and 
\begin{eqnarray}
G_{n\,\omega }^{N,R}(x,x^{\prime } =0)-G_{n\,\omega }^{N,A}(x,x^{\prime
}=0)=&-{\frac{2\pi i}L}\sum_{\ell =0}^\infty {\frac{\cos (\frac{\ell \pi x}L)%
}{(1+\delta _{\ell 0})\omega _n^\ell }}  \nonumber \\
&\times \left( \delta (\omega -\omega _n^\ell )-\delta (\omega +\omega
_n^\ell )\right).  \label{gnn}
\end{eqnarray}
In general, there are no monochromatic solutions for the problem of moving
boundaries, and hence it is not possible to write a normal mode
decomposition for the field in this case. However, since ${{\bf A}_{{\rm in}}%
}_n^{{\rm \scriptscriptstyle TE}}$ and ${{\bf A}_{{\rm out}}}_n^{{\rm %
\scriptscriptstyle TE}}$ satisfy the boundary conditions for two standing
mirrors at $x=0$ and $x=L,$ we may write their normal mode decompositions as
in Eqs.~(\ref{nmte})--(\ref{nmtm}), in terms of input and output bosonic
operators ${a_{{\rm in}}}_n^{{\rm \scriptscriptstyle TE}}$ and ${a_{{\rm out}%
}}_n^{{\rm \scriptscriptstyle (TE)}}$ (at this point our method is quite
similar to the approach developed in Ref.~\cite{bb} for the problem of
time--dependent material coefficients).  We then take the Fourier transform
of Eq.~(\ref{bcte}) and replace the result, jointly with Eq.~(\ref{gdd}),
into Eq.~(\ref{doug1}) in order to find the linear transformation between
the input and output TE bosonic operators: 
\begin{eqnarray}
{a_{{\rm out}}}_{n\ell }^{{\rm \scriptscriptstyle (TE)}} &=&{a_{{\rm in}}}%
_{n\ell }^{{\rm \scriptscriptstyle (TE)}}+\frac iL\sum_{\ell ^{\prime }=1}{%
\frac{\frac{\ell \pi }L\frac{\ell ^{\prime }\pi }L}{\sqrt{\omega _n^\ell
\omega _n^{\ell ^{\prime }}}}}\Bigl[\delta \!q[\omega _n^\ell -\omega
_n^{\ell ^{\prime }}]{a_{{\rm in}}}_{n\ell ^{\prime }}^{{\rm %
\scriptscriptstyle (TE)}}  \nonumber \\
&&+\delta \!q[\omega _n^\ell +\omega _n^{\ell ^{\prime }}]({a_{{\rm in}}}%
_{-n\ell ^{\prime }}^{{\rm \scriptscriptstyle (TE)}})^{\dagger }\Bigr],
\label{doug3}
\end{eqnarray}
where $\delta \!q[\omega]$ is the Fourier transform of $\delta \!q(t).$
The relation between TM operators is derived from Eqs.~(\ref{bctm}), (\ref
{doug2}) and (\ref{gnn}) in a similar way: 
\begin{equation}
{a_{{\rm out}}}_{n\ell }^{{\rm \scriptscriptstyle (TM)}}={a_{{\rm in}}}%
_{n\ell }^{{\rm \scriptscriptstyle (TM)}}-\frac iL\sum_{\ell ^{\prime }=0}%
\Bigl[(1+\delta _{\ell 0})(1+\delta _{\ell ^{\prime }0})\Bigr]^{-{\frac 12}%
}\times  \label{doug4}
\end{equation}
\[
\times \Biggl\{ {\frac{(k_{\parallel }^n)^2-\omega _n^\ell \omega _n^{\ell
^{\prime }}}{\sqrt{\omega _n^\ell \omega _n^{\ell ^{\prime }}}}}\delta
\!q[\omega _n^\ell -\omega _n^{\ell ^{\prime }}]{a_{{\rm in}}}_{n\ell
^{\prime }}^{{\rm \scriptscriptstyle (TM)}}+{\frac{(k_{\parallel
}^n)^2+\omega _n^\ell \omega _n^{\ell ^{\prime }}}{\sqrt{\omega _n^\ell
\omega _n^{\ell ^{\prime }}}}}\delta \!q[\omega _n^\ell +\omega _n^{\ell
^{\prime }}]({a_{{\rm in}}}_{-n\ell ^{\prime }}^{{\rm \scriptscriptstyle (TM)%
}})^{\dagger }\Biggr\}. 
\]

From Eqs.~(\ref{doug3}) and (\ref{doug4}) we may readily derive the number
of photons generated inside the cavity as a quantum effect of the mirror's
motion. As discussed below, the effect is associated to the creation
operators appearing in the r.-h.-s. of Eqs.~(\ref{doug3}) and (\ref{doug4}).
We assume that the field is initially in the vacuum state. The motion of the
mirror then excites a given number of photons $N_{n,\ell }^j$ with indexes $%
n,\ell $ and polarization $j.$ $N_{n,\ell }^j$ is given by the corresponding
output number operator averaged over the input vacuum state: 
\begin{equation}
N_{n,\ell }^j=\langle 0,{\rm in}\mid ({a_{{\rm out}}}_{n\ell }^j)^{\dagger }{%
a_{{\rm out}}}_{n\ell }^j\mid 0,{\rm in}\rangle ,  \label{avera}
\end{equation}
Replacing Eqs.~(\ref{doug3}) and (\ref{doug4}) into (\ref{avera}) provides
the photon numbers for each polarization: 
\begin{equation}
N_{n,{\ell }}^{{\rm \scriptscriptstyle (TE)}}=\frac 1{L^2}\sum_{\ell
^{\prime }=1}\left( \frac{\ell \pi }L\right) ^2\left( \frac{\ell ^{\prime
}\pi }L\right) ^2\frac 1{\omega _n^\ell \omega _n^{{\ell }^{\prime }}}\mid
\delta \!q[\omega _n^\ell +\omega _n^{{\ell }^{\prime }}]\mid ^2,
\label{res1te}
\end{equation}
and 
\begin{equation}
N_{n,{\ell }}^{{\rm \scriptscriptstyle (TM)}}=\frac 1{L^2}\sum_{{\ell }%
^{\prime }=0}{\frac{\left( (k_{\parallel }^n)^2+\omega _n^\ell \omega _n^{{%
\ell }^{\prime }}\right) ^2}{(1+\delta _{\ell 0})(1+\delta _{\ell ^{\prime
}0})\omega _n^\ell \omega _n^{{\ell }^{\prime }}}}\mid \delta \!q[\omega
_n^\ell +\omega _n^{{\ell }^{\prime }}]\mid ^2.  \label{res1tm}
\end{equation}
Since the frequencies $\omega _n^{\ell ^{\prime }}$ are positive, we infer
from Eqs.~(\ref{res1te}) and (\ref{res1tm}) that a given mechanical
frequency $\omega _0$ generates photons with frequencies $\omega _n^\ell
\leq \omega _0,$ thereby justifying the ansatz employed in this section.

From the above results we may directly calculate the photon production rates
and then estimate the order of magnitude of the quantum radiation effect.
Before addressing this question, however, we present a second derivation of
Eqs.~(\ref{res1te}) and (\ref{res1tm}), which is based on usual
time--dependent Hamiltonian perturbation theory. Note that the invariance of
the r.-h.-s. of Eqs.~(\ref{res1te}) and (\ref{res1tm}) with respect to the
permutation of $\ell $ and $\ell ^{\prime }$ suggests that the photons are
emitted in pairs. That this is indeed the case is more clearly shown by this
alternative approach, to be presented in the next section.

\section{Connection with radiation pressure}

Rather than considering the boundary conditions of a moving mirror, we
follow in this section the heuristic approach, first presented in Ref.~\cite
{jr1}, in which the effect of the mirror's motion is modelled by taking the
perturbation Hamiltonian 
\begin{equation}
\delta\!H=-\delta \!q(t)F,  \label{deltah}
\end{equation}
where $F$ is the field quantum operator representing the force on the moving
mirror. Accordingly, $\delta\!H$ corresponds to the energy supplied to the
field by means of the vacuum radiation pressure effect. The total
Hamiltonian of the field is 
\begin{equation}
H=H^{(0)}+\delta\!H,
\end{equation}
where the unperturbed Hamiltonian $H^{(0)}$ is written in terms of the
bosonic field operators for a standing mirror (see Eqs.~(\ref{nmte}) and (%
\ref{nmtm})) as 
\begin{equation}
H^{(0)}=\sum_{n,\ell }\sum_{j={\rm TE,TM}}\hbar \omega _n^\ell [(a_{n\ell
}^j)^{\dagger }a_{n\ell }^j+1/2].
\end{equation}
As discussed elsewhere~\cite{moore}, a Hamiltonian approach is not
rigorously consistent with the model of perfect reflectiveness considered
here. However, this model may be considered as an approximation for
dielectric mirrors with large refraction index $n$ --- for which a rigorous
Hamiltonian model is available~\cite{barton}, although such correspondence
is not yet settled (according to Ref.~\cite{barside2}, some unexpected
results show up when taking the limit of large $n$). In any case, the
formalism presented in this section is justified by comparing the results it
provides with those obtained in Sec.~II.

The force operator is the integral over the surface of the mirror (at its
rest position at $x=0$) of the $xx$ component of the Maxwell stress tensor: 
\begin{equation}
F={\frac 12}\int d^2{\bf r}_{\parallel }(E_x(0^{+})^2-B_{\parallel
}(0^{+})^2),  \label{force}
\end{equation}
where the limit $x\rightarrow 0$ is taken from positive values of $x$ as
indicated above (as in the previous section, we do not analyze the effect of
the field outside the plane cavity). Since $F$ is a quadratic operator on
the field, the perturbation Hamiltonian $\delta\!H$ excites pairs of photons
like in the problem of parametric amplification by a $\chi ^{(2)}$ nonlinear
medium. Thus, we consider a perturbed field state of the form 
\begin{equation}
|\Psi \rangle =\sum_{\{n\ell j,n^{\prime }\ell ^{\prime }j^{\prime
}\}}c_{\{n\ell j,n^{\prime }\ell ^{\prime }j^{\prime }\}}(t)|\{n\ell
j,n^{\prime }\ell ^{\prime }j^{\prime }\}\rangle +b(t)|0\rangle ,
\label{psi}
\end{equation}
where we sum over all two--photon states $|\{n\ell j,n^{\prime }\ell
^{\prime }j^{\prime }\}\rangle $ (the symbols $j$ and $j^{\prime }$
representing the polarizations of the photons in a given pair $\{n\ell
j,n^{\prime }\ell ^{\prime }j^{\prime }\}$). Note that each pair $\{n\ell
j,n^{\prime }\ell ^{\prime }j^{\prime }\}$ is included only once in Eq.~(\ref
{psi}), regardless of the ordering of the indices.

We assume that at $t\rightarrow -\infty$ the field is in the vacuum state,
so that the two--photon amplitudes are initially zero: $c_{\{n\ell
j,n^{\prime}\ell^{\prime}j^{\prime}\}}(-\infty)=0, b(-\infty)=1.$ We compute
the build--up of the two--photon amplitude $c_{\{n\ell
j,n^{\prime}\ell^{\prime}j^{\prime}\}}(t)$ from standard first--order
perturbation theory: 
\begin{equation}
c_{\{n\ell j,n^{\prime}\ell^{\prime}j^{\prime}\}}(t)= -{\frac{i}{\hbar}}%
\int_{-\infty}^t \langle n\ell j,n^{\prime}\ell^{\prime}j^{\prime}|
\delta\!H(t^{\prime})|0\rangle \exp\Bigl[{\frac{i}{\hbar}}(E^{(0)}_{n\ell,n^{\prime}%
\ell^{\prime}}- E^{(0)}_{{\rm vac}})t^{\prime}\Bigr]dt^{\prime},
\label{standard}
\end{equation}
with 
\begin{equation}
E^{(0)}_{n\ell,n^{\prime}\ell^{\prime}}-E^{(0)}_{{\rm vac}%
}=\hbar(\omega_n^{\ell}+\omega_{n^{\prime}} ^{\ell^{\prime}})  \label{endif}
\end{equation}
representing the difference between the (unperturbed) energies of the final
and initial states. As discussed in the previous sections, it is meaningless
to discuss two--photon amplitudes as long as the mirror is moving.
Accordingly, we must take $t\rightarrow\infty$ in Eq.~(\ref{standard}) in
order to have a consistent picture of the quantum radiation effect. Then,
replacing Eqs.~(\ref{deltah}) and (\ref{endif}) into Eq.~(\ref{standard})
yields 
\begin{equation}
c_{\{n\ell j,n^{\prime}\ell^{\prime}j^{\prime}\}}(\infty)={\frac{i}{\hbar}}
\langle n\ell
j,n^{\prime}\ell^{\prime}j^{\prime}|F|0\rangle\delta\!q[\omega_n^{\ell}+%
\omega_{n^{\prime}}^{\ell^{\prime}}].  \label{pertur1}
\end{equation}

In order to compute the matrix element appearing in the r.-h.-s. of Eq.~(\ref
{pertur1}), we write the electric and magnetic fields in Eq.~(\ref{force})
in terms of the potentials ${\bf A}^{{\rm \scriptscriptstyle (TE)}}$ and $%
\mbox{\boldmath$\cal A$}^{{\rm \scriptscriptstyle
(TM)}}.$ It is convenient to use the Fourier series representation defined
by 
\begin{equation}
{\bf A}^{{\rm \scriptscriptstyle (TE)}}(x,{\bf r}_{\parallel},t)= \sum_n 
{\bf A}_n^{{\rm \scriptscriptstyle (TE)}}(x,t) \exp(i {\bf k}_{\parallel}^n
\cdot {\bf r}_{\parallel }),  \label{fs}
\end{equation}
and by an equivalent expression for the TM potential $%
\mbox{\boldmath$\cal
A$}^{{\rm \scriptscriptstyle
(TM)}}.$ Then, the force operator is written as 
\begin{eqnarray}
F={\frac{S}{2}} \sum_n\Bigl[(k_{\parallel}^n )^2 \mbox{\boldmath$\cal A$}_n
^{{\rm \scriptscriptstyle (TM)}}(0^+,t)\cdot \mbox{\boldmath$\cal A$}_{-n} ^{%
{\rm \scriptscriptstyle (TM)}}(0^+,t)  \nonumber \\
-\partial_t \mbox{\boldmath$\cal A$}_n ^{{\rm \scriptscriptstyle (TM)}%
}(0^+,t)\cdot \partial_t \mbox{\boldmath$\cal A$}_{-n} ^{{\rm %
\scriptscriptstyle (TM)}}(0^+,t) - \partial_x {\bf A}_n^{{\rm %
\scriptscriptstyle (TE)}}(0^+,t)\cdot \partial_x {\bf A}_{-n}^{{\rm %
\scriptscriptstyle (TE)}}(0^+,t)\Bigr].  \label{force2}
\end{eqnarray}
From Eq.~(\ref{force2}), we obtain 
\begin{equation}
\langle n\ell\, {\rm \scriptstyle TE},n^{\prime}\ell^{\prime}\,{\rm %
\scriptstyle TM}|F|0\rangle=0.  \label{ftetm}
\end{equation}
Therefore, the photons belonging to a given emitted pair have the same
polarization. This is a general property of the plane symmetry of the
problem, rather than a consequence of the specific model considered in this
paper. Note however that it has been recently shown that TE-TM pairs may be
radiated in the case of lateral motion of the mirror~\cite{north}.

Using the normal mode decomposition of the field operators as given by Eqs.~(%
\ref{nmte}) and (\ref{nmtm}), we may calculate the TE--TE and TM--TM matrix
elements. We first find 
\begin{eqnarray}
\langle n\ell\, {\rm \scriptstyle TE},n^{\prime}\ell^{\prime}\,{\rm %
\scriptstyle TE}| \partial_x {\bf A}_N^{{\rm \scriptscriptstyle (TE)}%
}(0^+,t)\cdot \partial_x {\bf A}_{-N}^{{\rm \scriptscriptstyle (TE)}%
}(0^+,t^{\prime})|0\rangle=  \nonumber \\
{\frac{\hbar}{S L}} {\frac{{\frac{\ell\pi}{L}} {\frac{\ell^{\prime}\pi}{L}}}{%
\sqrt{\omega_n^{\ell} \omega_n^{\ell^{\prime}}}}} \Bigl(\delta_{n,N}%
\delta_{n^{\prime},-N}e^{i(\omega_n^{\ell}t+\omega_n^{\ell^{\prime}}t^{%
\prime})}
+\delta_{n,-N}\delta_{n^{\prime},N}e^{i(\omega_n^{\ell^{\prime}}t+\omega_n^{%
\ell}t^{\prime})}\Bigr),  \label{TETE}
\end{eqnarray}
and 
\begin{eqnarray}
\langle n\ell\, {\rm \scriptstyle TM},n^{\prime}\ell^{\prime}\,{\rm %
\scriptstyle TM}| \mbox{\boldmath$\cal A$}_N ^{{\rm \scriptscriptstyle (TM)}%
}(0^+,t)\cdot \mbox{\boldmath$\cal A$}_{-N} ^{{\rm \scriptscriptstyle (TM)}%
}(0^+,t^{\prime})|0\rangle= {\frac{\hbar}{S L}} (1+\delta_{\ell
0})(1+\delta_{\ell^{\prime}0})  \nonumber \\
\times\omega_n^{\ell}
\omega_n^{\ell^{\prime}}\Bigl(\delta_{n,N}\delta_{n^{\prime},-N}e^{i(%
\omega_n^{\ell}t+\omega_n^{\ell^{\prime}}t^{\prime})}
+\delta_{n,-N}\delta_{n^{\prime},N}e^{i(\omega_n^{\ell^{\prime}}t+\omega_n^{%
\ell}t^{\prime})}\Bigr),  \label{TMTM}
\end{eqnarray}
where use was made of the property $\omega_n^{\ell}=\omega_{-n}^{\ell}.$
Combining Eqs.~(\ref{TETE}) and (\ref{TMTM}) with Eq.~(\ref{force2}) leads
to 
\begin{equation}
\langle n\ell\, {\rm \scriptstyle TE},n^{\prime}\ell^{\prime}\,{\rm %
\scriptstyle TE}|F|0\rangle= -{\frac{\hbar}{L}} {\frac{{\frac{\ell\pi}{L}} {%
\frac{\ell^{\prime}\pi}{L}}}{\sqrt{\omega_n^{\ell} \omega_n^{\ell^{\prime}}}}%
} \delta_{n,-n^{\prime}},  \label{fte}
\end{equation}
and 
\begin{equation}
\langle n\ell\, {\rm \scriptstyle TM},n^{\prime}\ell^{\prime}\,{\rm %
\scriptstyle TM}| F|0\rangle= {\frac{\hbar}{L}} {\frac{(k_{\parallel
}^n)^2+\omega _n^{\ell} \omega_n^{{\ell}^{\prime }}}{(1+\delta_{\ell
0})(1+\delta_{\ell^{\prime}0})\sqrt{\omega _n^{\ell} \omega _n^{{\ell}%
^{\prime }}}}} \delta_{n,-n^{\prime}}.  \label{ftm}
\end{equation}

From Eqs.~(\ref{fte}) and (\ref{ftm}), we may immediately calculate the
amplitudes of creation of pairs of photons by combining them with Eq.~(\ref
{pertur1}). Here we write the results obtained in Appendix B for the more
general case where both mirrors are moving, so that the first mirror is at $%
x=\delta\!q_1(t)$ and the second mirror at $x=L + \delta\!q_2(t).$ The
resulting creation probabilities are: 
\begin{eqnarray}
|c_{\{n\ell {\rm \scriptscriptstyle TE}, n^{\prime}\ell^{\prime}{\rm %
\scriptscriptstyle TE}\}}|^2&=& \frac 1{L^2} \left( \frac{\ell\pi } L\right)
^2 \left( \frac{\ell^{\prime}\pi }L\right) ^2 \frac 1{\omega _n^{\ell}\omega
_n^{{\ell}^{\prime }}}\times  \label{probate} \\
& & \mid \delta\!q_1[\omega _n^{\ell}+\omega _n^{{\ell}^{\prime }}] -
(-1)^{\ell+\ell^{\prime}} \delta\!q_2[\omega _n^{\ell}+\omega _n^{{\ell}%
^{\prime }}] \mid ^2 \delta_{n,-n^{\prime}}  \nonumber
\end{eqnarray}
and 
\begin{eqnarray}
|c_{\{n\ell {\rm \scriptscriptstyle TM}, n^{\prime}\ell^{\prime}{\rm %
\scriptscriptstyle TM}\}}|^2 & = & \frac 1{L^2} {\frac{\left( (K_{\parallel
}^n)^2+\omega _n^{\ell} \omega_n^{{\ell}^{\prime }}\right) ^2 }{%
(1+\delta_{\ell 0})(1+\delta_{\ell^{\prime}0}) \omega _n^{\ell} \omega _n^{{%
\ell}^{\prime }}}}\times  \label{probatm} \\
& & \mid \delta\!q_1[\omega _n^{\ell}+\omega _n^{{\ell}^{\prime }}] -
(-1)^{\ell+\ell^{\prime}} \delta\!q_2[\omega _n^{\ell}+\omega _n^{{\ell}%
^{\prime }}] \mid^2 \delta_{n,-n^{\prime}}.  \nonumber
\end{eqnarray}
Note that the photons in a given pair have opposite values of ${\bf k}%
_{\parallel}^n,$ which is again a consequence of the plane symmetry~\cite
{pa4}. As shown in Appendix C, Eqs.~(\ref{TETE})--(\ref{probatm}) must be
slightly modified when considering the particular value $n=n^{\prime}=0$
(which corresponds to the 1D limit of our 3D formalism, since such modes
propagate along the $x$ direction and do not contain any dependence on the
transverse coordinates $y$ and $z$).

According to Eqs.~(\ref{probate}) and (\ref{probatm}), the joint motion of
the two mirrors selects the longitudinal modes according to the parity of
the indices $\ell .$ When $\delta \!q_1=-\delta \!q_2,$ which corresponds to
the `elongation mode' of the cavity, the two photons in a pair correspond to 
$\ell $ values of the same parity, the opposite taking place when the motion
is such that the cavity length is kept constant ($\delta \!q_1=\delta \!q_2$%
). This property is a straightforward generalization of the situation found
in one--dimensional cavities~\cite{astrid}. It shows that the radiation
effect should not be interpreted simply as a consequence of changing the
optical cavity length, since it also takes place when there is no relative
motion of the mirrors.

We may compute the average number of photons in a given cavity mode from 
\begin{equation}
N_{n,{\ell}}^{j}=\langle \Psi | (a_{n\ell}^{j})^{\dagger}a_{n\ell}^{j} |
\Psi \rangle.  \label{av}
\end{equation}
Replacing Eq.~(\ref{psi}) into Eq.~(\ref{av}) yields 
\begin{equation}
N_{n,{\ell}}^{j}=\sum_{\ell^{\prime}}|c_{\{n\ell j,-n\ell^{\prime}j\}}|^2.
\label{av2}
\end{equation}
Eqs.~(\ref{probate}) and (\ref{probatm}), in the particular case of $%
\delta\!q_2=0,$ jointly with Eq.~(\ref{av2}) provide results for the photon
numbers in full agreement with Eqs.~(\ref{res1te}) and (\ref{res1tm}) of the
previous section. As for the particular case with $n=0,$ Eq.~(\ref{av2})
also needs some slight modification in order to include the contribution of
the degenerate two--photon states $|0\ell,0\ell\rangle.$ As shown in
Appendix C, there is agreement with the results found in section 2 in this
case as well. We then conclude that the heuristic approach developed in this
section yields the same final expressions for the number of photons produced
in a given cavity mode. Moreover, it explicitly shows that the photons are
generated in pairs, the photons in a pair having the same polarization and
opposite values of ${{\bf k}_{\parallel}}_n.$

With the aid of the linear response formalism~\cite{kubo}, the perturbation
Hamiltonian as given by Eq.~(\ref{deltah}) may be also applied to compute
the dissipative part of the radiation pressure force $\delta F$ exerted on
the moving mirrors~\cite{jr1}--\cite{motion}. Such dissipative force is the
mechanical effect of the quantum radiation process, and hence must be
interpreted as a radiation reaction force. Since it generalizes Casimir's
result for a situation where (at least) one of the mirrors is moving, it has
been called motional Casimir force in Ref.~\cite{motion}, where a
one--dimensional calculation is presented for the case of
partially--transmitting mirrors. We consider again the case where one of the
mirrors is at rest, and then write the Fourier transformed force $\delta
F[\omega ]$ as: 
\begin{equation}
\delta F[\omega ]=\chi [\omega ]\delta \!q[\omega ].  \label{lin1}
\end{equation}
As discussed in Ref.~\cite{jr1}, linear response theory provides a result
for the imaginary part of the susceptibility function $\chi [\omega ],$
which corresponds to the dissipative component of the force, in terms of the
function $C_{FF}[\omega ]$ representing the spectrum of fluctuations of the
force operator on a standing mirror: 
\begin{equation}
{\rm Im}\chi [\omega ]={\frac 1{2\hbar }}(C_{FF}[\omega ]-C_{FF}[-\omega ]).
\label{lin2}
\end{equation}
The spectrum of fluctuations $C_{FF}[\omega ]$ is defined as the Fourier
transform of the force correlation function. It may be written in terms of
the two--photon matrix elements obtained above as follows~\cite{pa5}: 
\begin{equation}
C_{FF}[\omega ]=2\pi \sum_{\{n\ell j,n^{\prime }\ell ^{\prime }j^{\prime
}\}}\delta (\omega -\omega _n^\ell -\omega _{n^{\prime }}^{\ell ^{\prime
}})|\langle n\ell j,n^{\prime }\ell ^{\prime }j^{\prime }|F|0\rangle |^2.
\label{formula}
\end{equation}
where, as in Eq.~(\ref{psi}), each pair $\{n\ell j,n^{\prime }\ell ^{\prime
}j^{\prime }\}$ is included only once (regardless of the ordering).

The matrix elements of the force being given by Eqs.~(\ref{ftetm}), (\ref
{fte}) and (\ref{ftm}), we replace the r.-h.-s. of Eq.~(\ref{formula}) into
Eq.~(\ref{lin2}) to find 
\begin{eqnarray}
{\rm Im}\chi[\omega]={\frac{\pi\hbar}{2L^2}}\sum_{n,\ell,\ell^{\prime}}{%
\frac{({\frac{\ell\pi}{L}})^2 ({\frac{\ell^{\prime}\pi}{L}})^2+\left(
(k_{\parallel }^n)^2+\omega _n^{\ell} \omega_n^{{\ell}^{\prime }}\right) ^2 
}{(1+\delta_{\ell 0})(1+\delta_{\ell^{\prime}0})\omega _n^{\ell} \omega _n^{{%
\ell}^{\prime }}}}  \nonumber \\
\times\Bigl[\delta(\omega-\omega_n^{\ell}-\omega_n^{\ell^{\prime}}) -
\delta(\omega+ \omega_n^{\ell}+\omega_n^{\ell^{\prime}})\Bigr].
\label{force3}
\end{eqnarray}
Eq.~(\ref{force3}) provides the result for the dissipative component of the
force exerted on the mirror. The term with $n=0$ in Eq.~(\ref{force3}) is
particularly interesting because it allows for a comparison with the results
obtained in Ref.~\cite{motion} for a 1D scalar field. As discussed in
Appendix C, we find that in this case the two polarizations (represented by
the two terms in the r.-h.-s. of Eq.~(\ref{force3})) give identical
contributions to the dissipative susceptibility, which are in agreement with
the perfectly--reflecting limit of the 1D susceptibility function derived in
Ref.~\cite{motion}.

As mentioned before, ${\rm Im}\chi[\omega]$ is directly related to the
number of radiated photons by energy conservation. Indeed, comparing Eqs.~(%
\ref{res1te}) and (\ref{res1tm}) with Eq.~(\ref{force3}), we find 
\begin{equation}
\sum_{n,\ell}\hbar \omega_n^{\ell} (N_{n,\ell}^{{\rm \scriptscriptstyle (TE)}%
}+ N_{n,\ell}^{{\rm \scriptscriptstyle (TM)}})=\int{\frac{d\omega}{2\pi}}
\omega({\rm Im}\chi[\omega])|\delta\!q[\omega]|^2.  \label{comp1}
\end{equation}
Eq.~(\ref{comp1}) shows that the energy supplied to the field by the
radiation pressure force $\delta F[\omega],$ given by its r.-h.-s., is equal
to the total radiated energy. In the next section, we discuss in detail the
properties of the radiation by taking the specific example of sinusoidal
motion.

\section{Photon production rates}

In this section, we discuss in some detail the properties of the radiation
emitted inside the cavity, starting from the expressions for the two--photon
probabilities given by Eqs.~(\ref{probate}) and (\ref{probatm}), which were
shown to agree with the results for the photon numbers $N_{n,\ell}$ obtained
directly from the moving boundary conditions and given by Eqs.~(\ref{res1te}%
) and (\ref{res1tm}). We assume that the second cavity mirror is at rest at $%
x=L$ (hence $\delta\!q_2=0$), and that the first mirror oscillates around $%
x=0$ according to the law: 
\begin{equation}
\delta\!q(t)=\delta\!q_0 e^{-\mid t\mid /\Delta\! t} \cos(\omega_0 t),
\label{doug5}
\end{equation}
where the amplitude $\delta q_0$ and frequency $\omega_0$ satisfy the
non-relativistic condition $\omega_0\delta\!q_0 \ll 1.$ Moreover, we assume
that the damping time $\Delta\! t$ is much larger than the period of the
mechanical oscillation: 
\[
\omega_0\, \Delta\! t\gg 1. 
\]

We first consider the 1D limit of the results found in sections 2 and 3, by
picking up the photon pairs with $n=0.$ Refs.~\cite{Law} and \cite{dodo}
presented a 1D nonperturbative treatment for the situation where the
mechanical frequency $\omega _0$ satisfies the resonance condition 
\begin{equation}
\omega _0={\frac{\pi (\ell +\ell ^{\prime })}L},  \label{reson}
\end{equation}
for two longitudinal cavity modes $\ell $ and $\ell ^{\prime }$ (Ref.~\cite
{dodo} considered the particular case $\ell =\ell ^{\prime }=1,$ whereas
Ref.~\cite{Law} also considered the case $\ell =2,\ell ^{\prime }=1$). We
may discuss the relation between such formalism and the one presented in
this paper by taking the Fourier transform of Eq.~(\ref{doug5}) and
computing the two--photon probabilities in the resonant case (we omit
explicit reference to polarization while discussing the 1D limit). As shown
in Appendix C, we find 
\begin{equation}
|c_{\{0\ell ,0\ell ^{\prime }\}}|^2={\frac{\pi ^2\ell \ell ^{\prime }}{%
(1+\delta _{\ell \ell ^{\prime }})L^2}}(\delta \!q_0)^2\Delta \!t^2.
\label{1d2}
\end{equation}
According to Eqs.~(\ref{av2}) and (\ref{1d2}), the number of photons grows
quadratically in time in this case. The same time dependence may be obtained
as the short time limit of the 1D nonperturbative results found in Refs.~%
\cite{Law} and \cite{dodo}. Such behavior is related to the property that
the spectrum of a 1D perfect cavity is discrete, and it was also obtained in
the model of a 3D perfect closed cavity system discussed in Ref.~\cite{dodo}%
. In the case of a continuous spectrum, on the other hand, the emission
probabilities grow linearly in time as long as the perturbative
approximation is valid, which is well--known from the derivation of Fermi's
golden rule, so that in the end the meaningful physical quantities are the
photon production {\it rates,} as we show below. That is the case of a
partially--transmitting cavity, even in the 1D approximation (see Ref.~\cite
{astrid}), as well as of a 3D open cavity, as for instance the two parallel
infinite plates considered in this paper, even under the assumption
(considered in this paper) of perfect reflectiveness.

In the 3D case, we have to sum over all possible values of ${{\bf k}%
_{\parallel }}_n$ in order to compute the probability $\delta \!{\cal P}%
_{\ell _1,\ell _2}^j$ for emission of a pair of photons with indices $\ell
_1 $ and $\ell _2$ and polarization $j.$ Since ${{\bf k}_{\parallel }}_n$ is
actually a continuous variable, we replace 
\[
\sum_n={\frac S{(2\pi )^2}}\int d^2{k}_{\parallel }. 
\]
The probabilities do not dependent on the direction of ${{\bf k}_{\parallel }%
}_n,$ hence we find, first for TE polarization, 
\begin{equation}
\delta \!{\cal P}_{\ell _1,\ell _2}^{{\rm \scriptscriptstyle TE}}={\frac S{%
2\pi }}\int_0^\infty d\omega \,\omega |c_{\{\ell _1{\rm \scriptscriptstyle TE%
},\ell _2{\rm \scriptscriptstyle TE}\}}(\omega )|^2,  \label{int4}
\end{equation}
where $|c_{\{\ell _1{\rm \scriptscriptstyle TE},\ell _2{\rm %
\scriptscriptstyle TE}\}}(\omega )|^2$ is obtained from Eq.~(\ref{probate}): 
\begin{equation}
|c_{\{\ell _1{\rm \scriptscriptstyle TE},\ell _2{\rm \scriptscriptstyle TE}%
\}}(\omega )|^2=\frac 1{L^2}\left( \frac{\ell _1\pi }L\right) ^2\left( \frac{%
\ell _2\pi }L\right) ^2{\frac{|\delta \!q[\omega +\tilde{\omega}_{\ell
_1\ell _2}]|^2}{\omega \tilde{\omega}_{\ell _1\ell _2}}},  \label{c4}
\end{equation}
where 
\[
\tilde{\omega}_{\ell _1\ell _2}=\sqrt{\omega ^2-\left( \frac{\ell _1\pi }L%
\right) ^2+\left( \frac{\ell _2\pi }L\right) ^2} 
\]
represents the frequency of the `twin' photon of index $\ell _2$ and which
is emitted simultaneously with the photon of frequency $\omega $ and index $%
\ell _1.$ We perform the integral in Eq.~(\ref{int4}) in the limit of very
large $\Delta \!t,$ so that $\delta \!q[\omega ]$ is sharply peaked around $%
\omega _0.$ In this case, each pair $\ell _1,\ell _2$ determines completely
the frequencies $\omega _1$ and $\omega _2$ of the two photons. Moreover, it
also implies well defined values for the angles between the direction of
emission and the $x$ direction, which we denote as $\theta _1$ and $\theta
_2.$ In fact, we have 
\begin{equation}
\omega _1+\omega _2=\omega _0,  \label{set1}
\end{equation}
as in the problem of parametric amplification by a $\chi ^{(2)}$ nonlinear
medium; 
\begin{equation}
\omega _1\sin \theta _1=\omega _2\sin \theta _2  \label{set2}
\end{equation}
expressing the plane symmetry of the cavity, and which is loosely analogous
to the phase matching condition in nonlinear optics; and finally two
additional equations which result from the boundary conditions on the two
cavity mirrors: 
\begin{equation}
\omega _i\cos \theta _i={\frac{\ell _i\pi }L},  \label{set3}
\end{equation}
with $i=1,2.$ Eqs.~(\ref{set1})--(\ref{set3}) may be solved for $\omega _1,$ 
$\omega _2,$ $\theta _1$ and $\theta _2$ as functions of $\omega _0,$ $\ell
_1$ and $\ell _2.$ We find 
\begin{equation}
\omega _1={\frac{\omega _0}2}\left( 1+{\frac{\ell _1^2-\ell _2^2}{\beta ^2}}%
\right) ,  \label{resome}
\end{equation}
where 
\[
\beta =\omega _0L/\pi 
\]
is the ratio between the cavity round trip time--of--flight and the
mechanical period. Accordingly, the spectrum of photon emission, which is
continuous in the case of a single moving mirror~\cite{pa4}, becomes
discrete as a consequence of the two additional conditions, given by Eq.~(%
\ref{set3}), and which are associated to the presence of the second mirror
that constitute the cavity. For a given value of $\beta ,$ the set of
emitted frequencies is obtained from Eq.~(\ref{resome}) by taking all
positive integer values of $\ell _1$ and $\ell _2$ in the range defined by 
\begin{equation}
\ell _1+\ell _2\le \beta .  \label{range}
\end{equation}
In the case of TM polarization, the values $\ell _1=0$ and $\ell _2=0$ are
also allowed --- they correspond to travelling wave modes propagating
parallel to the plane of the mirror (waveguide modes). As for the spatial
direction of emission, the photons are emitted along directions defined by a
set of cones (whose axis of symmetry is the $x$--direction), each pair $\ell
_1,$ $\ell _2$ defining allowed values for $\theta _1$ and $\theta _2$
according to Eqs.~(\ref{set1})--(\ref{set3}). As an example, consider the
value $\beta =2\sqrt{2}.$ Since $\beta <3,$ the only allowed values for TE
polarization are $\ell _1=\ell _2=1,$ corresponding to a pair with $\omega
_1=\omega _2=\omega _0/2,$ and $\theta _1=\theta _2=45^o,$ which is however
not emitted in the case of rigid motion of the cavity. For TM polarization,
on the other hand, there are two additional pairs: one with $\ell _1=1,$ $%
\ell _2=0$ (rigid motion), giving $\omega _1=9\omega _0/16,$ $\omega
_2=7\omega _0/16,$ $\theta _1\approx 51^o,$ and $\theta _2=90^o;$ the other
with $\ell _1=\ell _2=0$ (elongation motion), giving $\omega _1=\omega
_2=\omega _0/2,$ and $\theta _1=\theta _2=90^o.$

We compute the photon production rate for emission at a given pair of
allowed frequencies assuming that the integrand in Eq.~(\ref{int4}) is the
product of a slowly--varying function of $\omega$ with the sharply--peaked
squared Fourier transform of $\delta\!q(t).$ This amounts to replace the
latter by a delta function, so that from Eq.~(\ref{doug5}) we derive: 
\begin{equation}
|\delta\!q[\omega+\tilde\omega_{\ell_1\ell_2}]|^2= {\frac{\pi}{2}}
(\delta\!q_0)^2 \Delta\!t {\frac{\omega_0-\omega_1}{\omega_0}}
\delta(\omega-\omega_1),  \label{deltaq}
\end{equation}
and noting that $\omega=\omega_1$ implies $\tilde\omega_{\ell_1\ell_2}=
\omega_2,$ we find the photon production rate of TE pairs with indices $%
\ell_1,\ell_2$ by replacing Eq.~(\ref{deltaq}) into Eq.~(\ref{c4}) and
performing the integral in Eq.~(\ref{int4}): 
\begin{equation}
W_{\ell_1,\ell_2}^{{\rm \scriptscriptstyle TE}}= {\frac{\delta\!{\cal P}^{%
{\rm \scriptscriptstyle TE}}_{\ell_1,\ell_2}}{\Delta\!t}}= {\frac{S}{4L^2}}%
\left( \frac{\ell_1\pi } L\right)^2 \left( \frac{\ell_2\pi }L\right)^2 {%
\frac{(\delta\!q_0)^2}{\omega_0}}.  \label{ratete}
\end{equation}
Note that the linear time dependence found for the probability $\delta\!%
{\cal P}^{{\rm \scriptscriptstyle TE}}_{\ell_1,\ell_2}$ originates from
integrating over the whole width of $|\delta\!q[\omega]|^2,$ instead of
taking just the peak value as in the derivation of the 1D result given by
Eq.~(\ref{1d2}). For TM photons, we find, starting from Eq.~(\ref{probatm})
and following the same method, 
\begin{equation}
W_{\ell_1,\ell_2}^{{\rm \scriptscriptstyle TM}}= {\frac{S}{4L^2}} {\frac{%
\left[\omega_0\omega_1-\left({\frac{\ell_1\pi}{L}}\right)^2\right]^2}{%
(1+\delta_{\ell_1 0})(1+\delta_{\ell_2 0})}} {\frac{(\delta\!q_0)^2}{\omega_0%
}}.  \label{ratetm}
\end{equation}

We may also derive the total production rate for a given value of $\beta$ by
adding over all values of $\ell_1$ and $\ell_2$ in the range defined by Eq.~(%
\ref{range}): 
\begin{equation}
W^j =\sum_{\ell_1,\ell_2}^{\ell_1+\ell_2\le \beta} W^j_{\ell_1,\ell_2},
\label{totrate}
\end{equation}
with $j={\rm TE}, {\rm TM}.$ In the figure, we plot $W^{{\rm %
\scriptscriptstyle TM}}$ and $W^{{\rm \scriptscriptstyle TE}},$ both divided
by the total production rate of TE photons in the case of a single moving
mirror (see Ref.~\cite{pa4}), 
\begin{equation}
W_{{\rm single}}^{{\rm \scriptscriptstyle TE}}= {\frac{1}{720 \pi^2}} S
(\delta\!q_0)^2\omega_0^5,  \label{single}
\end{equation}
as functions of $\beta.$ The curves displayed in the figure for TE and TM
polarizations are similar to those representing the decay rate of a
classical dipole at the midpoint between two perfect plane mirrors along the
direction parallel and perpendicular to the mirrors, respectively.
Underlying both effects are the properties of the vacuum field in the case
of a plane cavity geometry and the corresponding mode spectral density
function~\cite{haroche}. The most striking differences between the two
problems are related to the two--photon nature of the quantum radiation
effect considered in this paper (that explains, for instance, why, as
displayed in the figure, the TE photon production rate vanishes for $\beta <
2,$ whereas the parallel dipole decay rate vanishes for $\beta <1$ only).

As in the problem of a decaying dipole, the photon production rates jump at
integer values of $\beta .$ This originates from adding the contribution of
a new pair $\ell _1,\ell _2$ within the range defined by Eq.~(\ref{range}).
The jumps for TE polarization are comparatively larger, which may be
understood from the fact, discussed in detail in Ref.~\cite{pa4} in the case
of a single mirror, that TE photons are preferably emitted near the $x$
direction, thus being more sensitive to the discrete nature of the
wavevector along that direction. For both polarizations, the jumps become
smaller as $\beta $ increases, and then the curves approach their asymptotic
values for $\beta \rightarrow \infty $ which are indicated by the dashed
lines in the figure. As expected, they correspond to the photon production
rates for a single moving mirror --- the rate for TM polarization being $11$
times larger than the rate for TE polarization, given by Eq.~(\ref{single}).
Alternatively, the asymptotic limits may be derived directly from the
analytical results given by Eqs.~(\ref{ratete}) and (\ref{ratetm}) if we
replace the sum in Eq.~(\ref{totrate}) by an integral. In fact, performing
the integral in the case of TE polarization leads to the expression given by
Eq.~(\ref{single}), whereas the result for TM polarization comes with an
extra factor of $11.$

Of special interest is the behavior of the TM photon production rate in the
range $0<\beta <1,$ where, according to the figure, $W^{{\rm %
\scriptscriptstyle TM}}$ increases strongly as $\beta $ decreases to zero.
The precise dependence on $\beta $ may be obtained by replacing $\omega
_1=\omega _2=\omega _0/2$ in Eqs.~(\ref{ratetm}) and (\ref{totrate}) and
comparing with Eq.~(\ref{single}): 
\begin{equation}
W^{{\rm \scriptscriptstyle TM}}=W_{0,0}^{{\rm \scriptscriptstyle TM}}={\frac{%
45}4}{\frac{W_{{\rm single}}^{{\rm \scriptscriptstyle TE}}}{\beta ^2}}.
\label{menor1}
\end{equation}
Such dependence with $\beta $ suggests that the most favorable orders of
magnitude occur for $\beta <1.$ In this range, the photons have frequency $%
\omega _0/2$ and propagate along directions parallel to the mirrors.
Following Refs.~\cite{astrid} and \cite{pa4} we rewrite the photon
production rate given by Eq.~(\ref{menor1}) as: 
\begin{equation}
W^{{\rm \scriptscriptstyle TM}}={\frac 1{16}}{\frac S{\lambda _0^2}}\left( {%
\frac{v_{{\rm max}}}c}\right) ^2{\frac{\omega _0}{\beta ^2}},  \label{order1}
\end{equation}
where $v_{{\rm max}}=\omega _0\delta \!q_0$ is the maximum value of the
velocity, and $\lambda _0=2\pi c/\omega _0$ is half the value of the
wavelength of the emitted photons (we have reintroduced the speed of light $%
c $). As in Ref.~\cite{astrid}, we take $v_{{\rm max}}/c=10^{-7}$ and $%
\omega _0=2\pi \times 10^{10}{\rm sec}^{-1},$ yielding $\lambda _0=3{\rm cm}%
. $ A real experiment would hardly employ moving mirrors with transverse
dimensions larger than that, thus we take $S/\lambda _0^2\approx 1$ in order
to have a crude estimate of the orders of magnitude, even though diffraction
effects at the borders of the mirrors, not taken into account in this paper,
are of course relevant in this range. Finally, we take $L=1\mu {\rm m},$
giving $\beta \approx 10^{-4}.$ Eq.~(\ref{order1}) then yields $W^{{\rm %
\scriptscriptstyle TM}}\approx 4\times 10^3{\rm photons/sec}.$

\section{Conclusion}

We have calculated the photon production rates for a plane cavity with
moving mirrors by two different methods. In the first approach, we consider
the boundary conditions for perfectly reflecting moving mirrors in the long
wavelength approximation and assuming the field modification due to the
motion to be small. We then obtain an input--output transformation for the
field bosonic operators which allows us to compute the number of emitted
photons. In the second approach, we start from an effective perturbative
Hamiltonian  and apply usual first--order perturbation theory.
This method is considerably simpler since the expressions for the fields
scattered by a moving mirror are not required, and establishes a clear
connection between the photon emission effect and vacuum radiation pressure.
Furthermore, it explicitly unveils the fact that the photons are emitted in
pairs (that satisfy simple properties expressing the symmetry of the plane
geometry), essentially because the effect is contained in the time evolution
of the field state vectors rather than in the evolution of the field
operators.  The two methods provide the same results for the photon
production rates, hence justifying the somewhat heuristic Hamiltonian
approach.

Radiation is generated even when the distance between the mirrors is
kept constant, showing that the effect is not simply a  
consequence of modulating the optical cavity length. 
When the initial cavity length $L$ is much smaller than $2\pi c/\omega _0$
(we have considered in detail the example of a quasi--sinusoidal motion at
frequency $\omega _0$), however, radiation is emitted only in the case of
relative motion of the mirrors, and the generation rate is enhanced 
as $L$ decreases. 
In this regime, the photons are generated at the subharmonic frequency $%
\omega _0/2,$  propagate parallel to the plane of the mirror, and
are TM polarized. Such
enhancement effect is closely related to the properties of the radiation
emitted by a single mirror in free space \cite{pa4}, whose spectrum for TM
polarization is sharply peaked around the frequency $\omega _0/2.$

The orders of magnitude for the photon production rate found in this paper
suggest that the motion induced quantum radiation effect may be observed
under certain conditions. However, a careful analysis of the diffraction
effects near the border of the mirror would be necessary if a quantitative
comparison with experimental results is required, since the field
wavelengths involved would probably be of the order of the transverse
dimensions of the mirrors.

{\bf ACKNOWLEDGMENTS}\\This work was supported by Comisi\'{o}n Nacional de
Investigaci\'{o}n Cient\'{\i}fica y Tecnol\'{o}gica, Proyecto FONDECYT No.
2960009, and by Conselho Nacional de Desenvolvimento Cient\'{\i}fico e
Tecnol\'{o}gico.

\newpage

\appendix

\section{Boundary conditions for a perfect moving mirror}

In this appendix, we derive the boundary conditions in the case of a perfect
plane mirror moving along its normal direction. We take a Lorentz frame $%
S^{\prime}(t_0)$ whose trajectory in the laboratory frame $S$ is given by $%
x=\delta \!{\dot{q}(t_0)}(t-t_0)+\delta\!q(t_0),$ so that $S^{\prime}(t_0)$
represents the instantaneously co-moving frame at time $t_0.$ Quantities
measured in $S^{\prime}(t_0)$ are denoted by primed letters. The space--time
coordinates  in $S^{\prime}(t_0)$ are related to those in $S$ by 
\begin{equation}
x= \gamma\left(x^{\prime}+ \delta \!{\dot{q}(t_0)}t^{\prime}\right)+\delta%
\!q(t_0), {\bf r}^{\prime}_{\parallel}={\bf r}_{\parallel}, t=
\gamma\left(t^{\prime}+ \delta \!{\dot{q}(t_0)}x^{\prime}\right)+t_0,
\label{t1}
\end{equation}
where $\gamma=\left[1-(\delta \!{\dot{q}(t_0)})^2\right]^{-1/2}.$ The
electromagnetic
fields ${\bf E}^{\prime }$ and ${\bf B}^{\prime }$ satisfy the following
conditions: 
\begin{equation}
\widehat{{\bf x}}\times {\bf E}^{\prime }(x^{\prime}=0,{\bf r}%
^{\prime}_{\parallel }, t^{\prime}=0)=0; {\bf \ }\widehat{{\bf x}}\cdot {\bf %
B}^{\prime }(x^{\prime}=0,{\bf r}^{\prime}_{\parallel }, t^{\prime}=0)=0.
\label{ori}
\end{equation}
In the case of TE polarization, the condition for the electric field yields 
\begin{equation}
\partial_{t^{\prime}}{{\bf A}^{{\rm \scriptscriptstyle (TE)}}}%
^{\prime}(x^{\prime}=0,{\bf r}^{\prime}_{\parallel }, t^{\prime}=0)=0,
\label{cte1}
\end{equation}
and since $\hat{x}\cdot {{\bf A}^{{\rm \scriptscriptstyle (TE)}}}^{\prime}=0,
$ we have from Eqs.~(\ref{t1}) and (\ref{cte1}) 
\begin{equation}
\gamma\left( \delta \!{\dot{q}(t_0)}\partial_x +\partial_t\right) {\bf A}^{%
{\rm \scriptscriptstyle (TE)}}(x=\delta\!q(t_0),{\bf r}_{\parallel },t=t_0)=
\gamma d_t{\bf A}^{{\rm \scriptscriptstyle (TE)}}(x=\delta\!q(t=t_0),{\bf r}%
_{\parallel }, t=t_0)= 0.  \label{cte2}
\end{equation}
where $d_t$ represents the total time derivative. Since $t_0$ is arbitrary,
Eq.~(\ref{cte2}) implies that ${\bf A}^{{\rm \scriptscriptstyle (TE)}%
}(x=\delta\!q(t),{\bf r}_{\parallel }, t)$ must assume a constant value,
which is taken to be zero as in Eq.~(\ref{bc1}).

As for TM polarization, the condition on the electric field given by Eq.~(%
\ref{ori}) jointly with Eq.~(\ref{defitm}) yield 
\begin{equation}
\partial_{x^{\prime}}{\mbox{\boldmath$\cal
A$}^{{\rm \scriptscriptstyle (TM)}}}^{\prime}(x^{\prime}=0,{\bf r}%
^{\prime}_{\parallel }, t^{\prime}=0)=0,  \label{ctm1}
\end{equation}
On the other hand, we may write the l.-h.-s. of Eq.~(\ref{ctm1}) in terms of
unprimed quantities by  using again Eq.~(\ref{t1}) and the fact that $\hat{x}%
\cdot {\mbox{\boldmath$\cal
A$}^{{\rm \scriptscriptstyle (TM)}}}^{\prime}=0:$ 
\begin{equation}
\partial_{x^{\prime}}{\mbox{\boldmath$\cal
A$}^{{\rm \scriptscriptstyle (TM)}}}^{\prime}(x^{\prime}=0,{\bf r}%
^{\prime}_{\parallel }, t^{\prime}=0)= \gamma\left( \partial_x+\delta \!{%
\dot{q}(t_0)}\partial_t \right) {\mbox{\boldmath$\cal      \label{ctm2}
A$}^{{\rm \scriptscriptstyle (TM)}}}(x=\delta\!q(t_0), {\bf r}_{\parallel},
t=t_0),
\end{equation}
and then we obtain the boundary condition as given by Eq.~(\ref{bc2}) from
Eqs.~(\ref{ctm1}) and (\ref{ctm2}).

\section{Two moving mirrors}

In this Appendix, we consider the more general case where both mirrors move
along the $x$ direction. The first mirror is at $x=\delta q_{1}(t),$ whereas
the second one is at $x= L+\delta q_{2}(t).$ As before, $L$ represents the
initial cavity length. For TE polarization, the boundary condition at the
second mirror now reads 
\begin{equation}  \label{apdx1}
{\bf A}^{{\rm \scriptscriptstyle (TE)}}(L+\delta q_{2}(t),{\bf r}_{\|},t)=0,
\end{equation}
which yields, in the long wavelength and perturbative approximations, the
following additional boundary condition for the motion induced perturbation $%
{\bf \delta A}^{{\rm \scriptscriptstyle (TE)}}:$ 
\begin{eqnarray}  \label{apdx2}
{\bf \delta A}^{{\rm \scriptscriptstyle (TE)}}(L,{\bf r}_{\|},t) & = & -
\delta q_{2}(t) \partial_{x}{\bf A}_{{\rm sta}}^{{\rm \scriptscriptstyle (TE)%
}}(x=L,{\bf r}_{\|},t).
\end{eqnarray}
Working in the mixed reciprocal space and using the normal mode
decomposition of ${\bf A}_{{\rm sta}}^{{\rm \scriptscriptstyle (TE)}}$ as
given by Eq.~(\ref{nmte}), Eq. (\ref{apdx2}) leads to 
\begin{eqnarray}  \label{apdx5}
{\bf \delta A}_n^{{\rm \scriptscriptstyle (TE)}}(L,\omega) & = & -i \, {\
\sum_{\ell=1}^{\infty}} (-1)^{\ell} \left(\frac{\ell \pi}L\right)\sqrt{\frac{%
\hbar}{\omega_n^{\ell}SL}} \, \left(\,\delta q_{1}[\omega-\omega_n^{\ell}]%
{\it a}_{n \, \ell}^{{\rm \scriptscriptstyle (TE)}}\right.  \nonumber \\
& &\left. + \delta q_{1}[\omega+\omega_n^{\ell}]({\it a}_{{\small -}n \,
\ell}^{{\rm \scriptscriptstyle (TE)}} )^{\dag}\, \right) \, {\bf \hat{%
\epsilon}}_n.
\end{eqnarray}
Of particular interest in Eq.~(\ref{apdx5}) is the factor $\cos(\ell
\pi)=(-1)^{\ell}$ that comes from evaluating the $x$ derivative of ${\bf A}_{%
{\rm sta}}^{{\rm \scriptscriptstyle (TE)}}$ at $x=L.$ Eq.~(\ref{apdx5})
jointly with Eq.~(\ref{bcte}) define a boundary value problem to be solved
with the aid of the Green functions given by Eq.~(\ref{gd}). We first employ
the retarded Green function $G_{n\,\omega}^{D,R} (x,x^{\prime})$ to solve
for the total field ${\bf A}_n^{{\rm \scriptscriptstyle (TE)}}$ in terms of
the input field ${\bf A}_{{\rm in}, \, n}^{{\rm \scriptscriptstyle (TE)}}:$ 
\begin{eqnarray}
{\bf A}_n^{{\rm \scriptscriptstyle (TE)}}(x,\omega)={\bf A}_{{\rm in}, \,
n}^{{\rm \scriptscriptstyle (TE)}}(x,\omega) &+ {\bf \delta A}_n^{{\rm %
\scriptscriptstyle (TE)}}(L,\omega) \partial_{x^\prime}
G_n^{D,R}(x^{\prime}=L,x;\omega)  \nonumber \\
&- {\bf \delta A}_n^{{\rm \scriptscriptstyle (TE)}}(0,\omega)
\partial_{x^\prime} G_n^{D,R}(x^{\prime}=0,x;\omega).
\end{eqnarray}
As in Sec.~2, we also solve Eq.~(\ref{apdx5}) in terms of the output field $%
{\bf A}_{{\rm out}, \, n}^{{\rm \scriptscriptstyle (TE)}}$ with the aid of
the advanced Green function $G_{n\,\omega}^{D,A} (x,x^{\prime}).$ The
connection between output and input fields is then provided by the
difference 
\[
\partial_{x^\prime} G_{n\,\omega}^{D,R}(x^{\prime}=L,x) -\partial_{x^\prime}
G_{n\,\omega}^{D,A}(x^{\prime}=L,x) 
\]
\begin{equation}  \label{apdx7}
=-\frac{2 \pi i}L \, \sum_{\ell=1}^{\infty} (-1)^{\ell} \left( \frac{\ell\pi}%
L \right) \sin \left(\frac{\ell \pi x}L \right) \, \frac1{\omega_n^{\ell}}\,
\left(\delta (\omega-\omega_n^{\ell})-\delta (\omega+\omega_n^{\ell})\right).
\end{equation}

As explained in Sec.~2, we derive the linear transformation between output
and input TE bosonic operators from Eqs.~(\ref{apdx5})--(\ref{apdx7}) 
\[
a_{{\rm out}, \, n \, \ell}^{{\rm \scriptscriptstyle (TE)}}=a_{{\rm in}, \,
n\, \ell}^{{\rm \scriptscriptstyle (TE)}}- \frac{i}{L} \,
\sum_{\ell^{\prime}=1}^{\infty} \left( \frac{\ell \pi}L \right) \left( \frac{%
\ell^{\prime} \pi}L \right) \, \frac{1}{(\omega_n^{\ell}
\omega_n^{\ell^{\prime}})^{1/2}} 
\]
\[
\times\left\{ \left(\delta
q_{1}[\omega_n^{\ell}-\omega_n^{\ell^{\prime}}]-(-1)^{\ell+\ell^{\prime}}%
\delta q_{2}[\omega_n^{\ell}-\omega_n^{\ell^{\prime}}] \right) \, a_{{\rm in}%
, \,n \, \ell^{\prime}}^{{\rm \scriptscriptstyle (TE)}}+\right. 
\]
\begin{equation}
\left. \left( \delta
q_{1}[\omega_n^{\ell}+\omega_n^{\ell^{\prime}}]-(-1)^{\ell+\ell^{\prime}}%
\delta q_{2}[\omega_n^{\ell}+\omega_n^{\ell^{\prime}}]\right) \, (a_{{\rm in}%
, \, {\small -}n \, \ell^{\prime}}^{{\rm \scriptscriptstyle (TE)}%
})^{\dagger}\right\}.  \label{2te}
\end{equation}
From Eq.~(\ref{2te}) we may calculate the number of photons $N_{n,\ell}^{%
{\rm \scriptscriptstyle (TE)}}$ by taking the average of the output number
operator over the input vacuum state as in Eq.~(\ref{avera}). For TM
polarization, we extend the method employed in Sec.~2 to take into account
the motion of the second mirror exactly as discussed above for TE
polarization.

Alternatively, we may compute the photon numbers from the effective
perturbation Hamiltonian 
\begin{equation}
\delta H= -\delta\!q_1(t) F_1 -\delta\!q_2(t) F_2,  \label{2deltah}
\end{equation}
where $F_i$ is the force exerted on mirror $i$ by the vacuum field.
Following the procedure outlined in Sec.~3, we derive the two--photon
creation probabilities given by Eqs.~(\ref{probate}) and (\ref{probatm}). As
in the case of a single moving mirror, the results obtained through this
method are in full agreement with those obtained directly from the boundary
conditions.

\section{Photons emitted along the normal direction}

In this appendix, we consider in detail the contribution of the degenerate
two--photon states in the derivation of the photon numbers and of the
susceptibility function. First note that degenerate two--photon states
necessarily correspond to propagation along the direction perpendicular to
the plane of the mirror, i.e., they are of the form $|n=0\,\,\ell
\,\,j,n=0\,\,\ell \,\,j\rangle .$ The degenerate two--photon matrix elements
of the force operator are calculated from the representation of the force
operator in terms of the vector potentials, given by Eq.~(\ref{force2}), and
from the normal mode decompositions given by Eqs.~(\ref{nmte}) and (\ref
{nmtm}): 
\begin{equation}
\langle n=0\,\,\ell \,{\rm \scriptstyle TE},n=0\,\,\ell \,{\rm \scriptstyle %
TE}|F|0\rangle =-\langle n=0\,\,\ell \,{\rm \scriptstyle TM},n=0\,\,\ell \,%
{\rm \scriptstyle TM}|F|0\rangle =-{\frac{\ell \pi \hbar }{\sqrt{2}L^2}}.
\label{forap}
\end{equation}
These results are smaller than the values of the expressions given by Eqs.~(%
\ref{fte}) and (\ref{ftm}) at $n=0$ and $\ell =\ell ^{\prime }$ by a factor
of $\sqrt{2}.$ From them, we easily compute the degenerate two--photon
probabilities by using Eq.~(\ref{pertur1}), allowing us to write the correct
expression for $n=0:$ 
\begin{equation}
|c_{\{0\ell {\rm \scriptscriptstyle TE},0\ell ^{\prime }{\rm %
\scriptscriptstyle TE}\}}|^2=|c_{\{0\ell {\rm \scriptscriptstyle TM},0\ell
^{\prime }{\rm \scriptscriptstyle TM}\}}|^2=\frac 1{(1+\delta _{\ell \ell
^{\prime }})L^2}\left( \frac{\ell \pi }L\right) \left( \frac{\ell ^{\prime
}\pi }L\right) \mid \delta \!q[\frac{\ell \pi }L+\frac{\ell ^{\prime }\pi }L%
]\mid ^2.  \label{proba}
\end{equation}
According to Eq.~(\ref{proba}), the degenerate two--photon probabilities are
one--half the value found when replacing the values $n=0,$ $\ell =\ell
^{\prime },$ and $\delta \!q_2=0$ in Eqs.~(\ref{probate}) and (\ref{probatm}%
).

The contribution of degenerate two--photon states is found from Eq.~(\ref{av}%
): 
\begin{equation}
N_{0,{\ell }}^j=\sum_{\ell ^{\prime },\ell ^{\prime }\ne \ell }|c_{\{0\ell
j,0\ell ^{\prime }j\}}|^2+2|c_{\{0\ell j,0\ell j\}}|^2.  \label{ava1}
\end{equation}
The factor two multiplying the degenerate two--photon amplitude in the
r.-h.-s. of Eq.~(\ref{ava1}) cancels the additional factor one--half
appearing in Eq.~(\ref{proba}) for $\ell =\ell ^{\prime },$ then yielding a
result in full agreement with Eqs.~(\ref{res1te}) and (\ref{res1tm}). For
the specific example of motion given by Eq.~(\ref{doug5}), and assuming that
the mechanical frequency $\omega _0$ satisfies the resonant condition as
given by Eq.~(\ref{reson}), we derive from Eq.~(\ref{proba}) the expression
for the production rate of photons with $n=0$ given by Eq.~(\ref{1d2})

Since the results for the photon numbers are not modified when taking into
account the degenerate two--photon states, we expect that the formula for
the dissipative component of the susceptibility function, given by Eq.~(\ref
{force3}), should also be valid for $n=0,$ so as to preserve the connection
between dissipation and total radiated displayed by Eq.~(\ref{comp1}). In
fact, we may write separately the contribution of degenerate two--photon
states to the sum over pairs $\{n\ell ,n^{\prime }\ell ^{\prime }\}$ in Eq.~(%
\ref{formula}): 
\begin{eqnarray}
C_{FF}[\omega ] &=&\pi \sum_j\sum_{n,\ell ,\ell ^{\prime }}^{*}\delta
(\omega -\omega _n^\ell -\omega _n^{\ell ^{\prime }})|\langle n\ell j,n\ell
^{\prime }j|F|0\rangle |^2  \nonumber \\
&+&2\pi \sum_j\sum_\ell \delta (\omega -2\ell \pi /L)|\langle 0\ell j,0\ell
j|F|0\rangle |^2,  \label{fapp1}
\end{eqnarray}
where $\sum_{n,\ell ,\ell ^{\prime }}^{*}$ represents the sum over all
possible values of $n,$ $\ell $ and $\ell ^{\prime }$ excluding those where
simultaneously $n=0$ and $\ell =\ell ^{\prime }.$ As before, the factor
one--half found for the degenerate two--photon matrix element (given by Eq.~(%
\ref{proba})) is cancelled by the factor two appearing in the r.-h.-s. of
Eq.~(\ref{fapp1}). Hence we may write the expression for the 1D dissipative
susceptibility function, ${\rm Im}\chi _{\scriptstyle 1D}[\omega ],$ by
selecting directly from Eq.~(\ref{force3}) the terms with $n=0:$ 
\begin{equation}
{\rm Im}\chi _{\scriptstyle 1D}[\omega ]={\frac{\pi ^3\hbar }{L^4}}%
\sum_{\ell =1}^\infty \sum_{\ell ^{\prime }=1}^\infty \ell \ell ^{\prime
}\left[ \delta (\omega -(\ell +\ell ^{\prime })\pi /L)-\delta (\omega +(\ell
+\ell ^{\prime })\pi /L)\right] .  \label{1d}
\end{equation}
As for Ref.~\cite{motion}, the result for the (complete) susceptibility
function in the perfectly reflecting limit and in the particular case where
only one mirror moves reads 
\begin{equation}
\tilde{\chi}[\omega ]={\frac \hbar {6\pi }}\Bigl[{\frac{i\omega ^3}{%
1-e^{2i\omega L}}}+({\frac \pi L})^2(i\omega )\Bigl({\frac 12}-{\frac 1{%
1-e^{2i\omega L}}}\Bigr)\Bigr].  \label{jr1}
\end{equation}
In order to compare Eqs.~(\ref{1d}) and (\ref{jr1}), we must take the
imaginary part of $\tilde{\chi}[\omega ],$ then yielding, after some
algebra, 
\begin{equation}
{\rm Im}\tilde{\chi}[\omega ]={\frac \hbar {6\pi }}\Bigl[{\frac{\omega ^3}2}+%
{\frac 12}\left( {\frac \pi L}\right) ^4\sum_{n=-\infty }^\infty
n(n^2-1)\delta (\omega -n\pi /L)\Bigr].  \label{jr2}
\end{equation}
The first term in the r.-h.-s. of Eq.~(\ref{jr2}) represents the
contribution of the field outside the cavity, being equal to half the value
found for a two--sided single mirror in one--dimensional vacuum~\cite{ford}.
The second term, on the other hand, represents the contribution of the
intracavity field, which is, by inspection of Eqs.~(\ref{1d}) and (\ref{jr2}%
), equal to half the value found from taking the 1D limit in Eq.~(\ref
{force3}), the factor of two being related to the two polarizations taken
into account in the electromagnetic case.

\newpage

{\Large {\bf Figure Caption}}

\bigskip\bigskip\bigskip

Total production rates of TE (a) and TM (b) photons as functions of $%
\beta=\omega_0 L/\pi,$ which represents the ratio between the round--trip
time of flight and the mechanical period. The scale of the vertical axis is
such that the value one corresponds to the generation rate of TE photons for
a single moving mirror. The dashed lines provide the asymptotic limits for
large $\beta.$ They show that the photon production rates approach the
values corresponding to the case of a single moving mirror in this limit
(note that the single mirror TM photon generation rate is larger than the
single mirror TE rate by a factor of $11$).


\newpage

\begin{references}
\bibitem{schwinger}  J. Schwinger, Proc. Natl. Acad. Sci. USA {\bf 90}, 958
(1993); {\bf 90}, 2105 (1993); {\bf 90}, 4505 (1993), {\bf 90}, 7285 (1993); 
{\bf 91}, 6473 (1994).

\bibitem{astrid}  A. Lambrecht, M.-T. Jaekel and S. Reynaud, Phys. Rev.
Lett. {\bf 77}, 615 (1996).

\bibitem{barton}  G. Barton and C. Eberlein, Ann. Phys., N.Y., {\bf 227},
222 (1993).

\bibitem{ford}  L. H. Ford and A. Vilenkin, Phys. Rev. {\bf D 25}, 2569
(1982); P. A. Maia Neto and L. A. S. Machado, Brazilian J. Phys. {\bf 25},
324 (1995).

\bibitem{pa1}  P. A. Maia Neto, J. Phys. A: Math. Gen. {\bf 27}, 2167 (1994).

\bibitem{jr1}  M. T. Jaekel and S. Reynaud, Quantum Optics {\bf 4}, 39
(1992).

\bibitem{fusao}  V. B. Braginsky and F. Ya. Khalili, Phys. Lett. A {\bf 161}%
, 197 (1991); G. Barton {\it New aspects of the Casimir effect: fluctuations
and radiative reaction}, in {\it Cavity Quantum Electrodynamics}
(Supplement: Advances in Atomic, Molecular and Optical Physics), ed. P.
Berman (Academic Press, New York, 1993).

\bibitem{pa5}  P. A. Maia Neto and S. Reynaud, Phys. Rev. A {\bf 47} 1639
(1993).

\bibitem{motion}  M. T. Jaekel and S. Reynaud, J. Phys. I France {\bf 2},
149 (1992).

\bibitem{moore}  G. T. Moore, J. Math. Phys. {\bf 11}, 2679 (1970).

\bibitem{Law}  C. K. Law, Phys. Rev. A {\bf 49}, 433 (1994).

\bibitem{dodo}  V. V. Dodonov and A. B. Klimov, Phys. Rev. {\bf A53}, 2664
(1996).

\bibitem{barside2}  G. Barton and C. A. North, Ann. Phys. (NY) {\bf 252}, 72
(1996).

\bibitem{north}  C. A. North, {\it TE--TM quantum radiation in three
dimensions from a oscillating dieletric half-space}, pre-print 1997.

\bibitem{barside}  G. Barton, Ann. Phys. (NY) {\bf 245}, 361 (1996).

\bibitem{eberlein}  C. Eberlein, Phys. Rev. {\bf A 53}, 2772 (1996); Phys.
Rev. Lett. {\bf 76}, 3842 (1996).

\bibitem{bb}  Z. Bialynicka--Birula and I. Bialynicka--Birula, J. Optical
Soc. Am. {\bf B 4}, 1621 (1987).

\bibitem{pa4}  P. A. Maia Neto and L. A. S. Machado, Phys. Rev. {\bf A 54},
3420 (1996).

\bibitem{kubo}  R. Kubo, Rep. Progr. Phys. {\bf 29}, 255 (1966).

\bibitem{haroche}  S. Haroche, {\it Cavity quantum electrodynamics}, in {\it %
Fundamental Systems in Quantum Optics}, (Les Houches Summer School, Session
LIII), eds. J. Dalibard, J.-M. Raymond and J. Zinn--Justin (North--Holland,
Amsterdam, 1992).
\end{references}
\end{document}